\newcounter{myctr}

\documentclass{ws-acs}
\usepackage{url}
\usepackage{multirow}
\providecommand{\tabularnewline}{\\}
\begin{document}

\makeatletter
\def\@biblabel#1{[#1]}
\makeatother

\markboth{M. Kaan \"Ozt\"urk}{Dynamics of discrete opinions without compromise}

%%%%%%%%%%%%%%%%%%%%% Publisher's Area please ignore %%%%%%%%%%%%%%%
%
%\catchline{}{}{}{}{}
%
%%%%%%%%%%%%%%%%%%%%%%%%%%%%%%%%%%%%%%%%%%%%%%%%%%%%%%%%%%%%%%%%%%%%

\title{DYNAMICS OF DISCRETE OPINIONS WITHOUT COMPROMISE}

\author{\footnotesize M.~KAAN~\"OZT\"URK}

\address{Dept. of Information Systems and Technologies, Yeditepe University\\
34755 Ata\c{s}ehir, Istanbul,
Turkey\\
kaan.ozturk@yeditepe.edu.tr}

\maketitle

\begin{history}
\received{(received date)}
\revised{(revised date)}
%\accepted{(Day Month Year)}
%\comby{(xxxxxxxxxx)}
\end{history}

\begin{abstract}
A new agent-based, bounded-confidence model for discrete one-dimensional opinion dynamics is presented. The agents interact if their opinions do not differ more than a tolerance parameter. In pairwise interactions, one of the pair, randomly selected, converts to the opinion of the other. The model can be used to simulate cases where no compromise is possible, such as choices of substitute goods, or other exclusive choices. The homogeneous case with maximum tolerance is equivalent to the Gambler's Ruin problem. A homogeneous system always ends up in an absorbing state, which can have one or more surviving opinions. An upper bound for the final number of opinions is given. The distribution of absorption times fits the generalized extreme value distribution. The diffusion coefficient of an opinion increases linearly with the number of opinions within the tolerance parameter. A general master equation and specific Markov matrices are given. The software code developed for this study is provided as a supplement.
\end{abstract}


\section{Introduction}
\label{sec:intro}
Opinion dynamics, as the term is used in the physics community, studies the evolution of systems composed of a large number of individuals characterized by a state (the ``opinion'') and modify their state by repeated interactions whose rules are based on social observations rather than physical laws. The goal of opinion dynamics is to develop models that capture the essential behavior of social systems.

Several models for opinion dynamics are developed in the last decade~\cite{CasForLor2010}. Some of them are agent-based, containing a large number of simulated persons, called agents. Depending on the details of the model and on parameter values, agents may all converge to a single opinion (consensus), to two separate opinions (polarization) or to several opinions (fragmentation). Usual questions include whether agents always (or at all) reach a consensus, and under which conditions; how long it takes to reach an equilibrium state; what the effect of the topology of the interaction network is, etc.

In models where agents are chosen randomly in pairs, the updated opinion of each agent is a function of the interacting pair only. Examples include the Axelrod model~\cite{Axelrod1997}, the voter models~\cite{KrapivskyRedner2003,VazquezRedner2004}, the CODA model~\cite{Martins2008,MartinsEtAl2009}, the Deffuant-Weisbuch model~\cite{DeffuantEtAl2000,WeisbuchEtAl2002}, and their variants. The differences and similarities between these models and the present model are discussed in section~\ref{subsec:model-comparison}.
 
In the so-called ``bounded confidence'' models, two agents interact only if the difference between their opinions is less than a given threshold value, called ``uncertainty'', ``tolerance'', or ``confidence bound''. This constraint reflects the fact that people change their opinions only to a limited degree. Each person has a certain tolerance interval, and rejects to consider or discuss ideas that are outside this interval. Bounded confidence can be applied to either type of models discussed above, but if opinions are discrete, it makes sense only if there are three or more different opinions.

This paper is organized as follows: Section~\ref{subsec:model} introduces the model, and explains the rationale behind its assumptions. Section~\ref{subsec:model-comparison} compares the model with existing models that use pairwise interactions. Section~\ref{sec:dynamics} describes how agents move in opinion space with or without bias, and outlines the general properties of the absorbing (final) states. Section~\ref{sec:fully} analyzes fully-mixed (homogeneous) populations using Markov matrices and agent-based simulations. Section~\ref{sec:network} briefly discusses some features of the model under general network topologies. 

The C and Matlab codes of the programs used in this study are provided as a supplement to this paper. Those who wish to replicate the results can download the codes at \texttt{https://github.com/mkozturk/Opinion\_Dynamics}.

\subsection{The model of interaction}
\label{subsec:model}
The model deals with one-dimensional discrete opinions. Agents meet pairwise, and each agent converts the other one to its own opinion with some probability.

The model comprises $N$ agents, each carrying an integer opinion in the range $[1,Q]$. Two agents are chosen randomly, with opinions $q$ and $q^\prime$, respectively. They interact only if $1\leq|q-q^\prime|\leq d$, that is, if the opinions differ by no more than the confidence bound $d$, an integer between 1 and $Q-1$. If this condition holds, we call the pair of agents ``compatible''. In this study we assume that $d$ is the same for every individual.

In the more general case, agents can be taken as nodes on an acquaintance network. In this study (except for the last section) we assume that this network is a complete graph where everybody can interact with everybody.

At every time step of a simulation one pair of compatible agents is chosen randomly, so that at every step there is exactly one opinion change. Thus, ``time'' is defined as the number of conversions, not as the number of random meetings between agents.

Upon interaction, a random choice is made so that agents both adopt opinion $q$ with probability $p$ or opinion $q^\prime$ with probability $1-p$. In the symmetric (unbiased) case $p$ is set to 0.5.

A second random choice at the interaction stage may look superfluous, as agents are already chosen randomly. However, it allows us to set $p$ as a function of the chosen agents' properties. Various biases can be introduced that way.

\subsubsection*{Using a small number of opinions}
The number of opinions $Q$ can be set to any positive integer, but in this study we deliberately keep this value small, following the practice of social sciences. Questionnaires used in social research involve interval scales, such as the Likert scale, consisting of only five or seven (or occasionally nine) levels of responses~\cite{McBurneyWhite2010}. This is not only a convenient methodological choice, but also a reflection of cognitive limits. Studies indicate that respondents cannot distinguish further subdivisions of attitudes and opinions~\cite{Hulbert1975,Cox1980}. (This behavior may be related to the ``channel capacity'' of the human mind, expressed by Miller~\cite{Miller1956} as the ``seven plus or minus two rule''.) As a result of this limitation, consumers experience ``information overload'' when faced with more than ten options for a specific item, resulting in confusion about choices and dissatisfaction with their final choice~\cite{Malhotra1982}. These observations suggest that an opinion dynamics model must not have too many opinions, lest it disagree with human cognitive limits.

The same point is also raised by Urbig~\cite{Urbig2003}, who points out that an attitude and its communication are different things, and criticizes models that are ``\emph{based on the assumption that individuals can communicate the difference between an attitude of 0.5555 and 0.5556.}'' In Urbig's sophisticated model, agents have continuous opinions but when they interact only discrete values are communicated.

The number $Q$ of different categories (opinions) a person can distinguish on a given subject depends on the person's knowledge, experience, and familiarity with the subject. For simplicity, we assume that $Q$ is the same for all agents in this study. 

\subsubsection*{Applicability of the model} The model presented here applies to situations where a compromise is not possible or reasonable. The lack of compromise implies discreteness of opinions because the initial set of opinions in the population is closed under the described interaction.

One obvious application is to political opinions. When a pair of voters interact, one may convert the other to vote for his party, but their interaction almost never results in both of them voting for a third party in the middle.

Another application area is consumer behavior. Consider a set of substitute goods, such as different brands of coffee (or of shirts, pens, toothpaste, ...), ordered according to their price. After interaction, one agent may decide that the merits of his correspondent's more expensive good warrants the extra cost and adopt it (``This coffee tastes much better, and it costs only one more dollar.''). Conversely, the other agent may adopt the cheaper good after realizing that it meets her needs (``This coffee is cheaper, and it tastes about the same.''). In the case of downgrade, the confidence bound is set by the minimum quality required by the consumer, while in the upgrade it is set by the maximum price the consumer is willing to pay. An ordering in price will usually be an ordering in quality, too, so the two bounds are consistent.

The non-compromise model would be better suited to goods appealing to personal, non-quantifiable taste. With goods such as computers, cell phones, or service plans, quantifiable properties may lead to a more rational comparison where both agents may end up adopting a third choice.

If an agent represents a geographic area, a lattice of agents can be used to investigate spatial distribution of opinions and to simulate the spread (or decline) of a specific choice over a geographic region. The opinions may represent different brands of substitute goods, or armies engaged in hostilities. When two agents (areas) interact and one is converted to the opinion of the other, the brand or army associated with the winning opinion becomes dominant in both geographic areas.

The model is also applicable to opinions that cannot be ordered. For example, we may need to choose among policies A, B, etc., about a given subject. Then opinion A would be expressed as ``I believe choice A is the best''. In such cases the distance between two opinions may not be easily quantified, but, depending on the problem, one can use multidimensional comparison, common sense, or intuition to decide if they are within each other's tolerance limit.

Consider the following scenario which can be modelled with parameters $Q=3$ and $d=1$: In a metropolitan city, three policies are proposed to deal with traffic congestion. Policy 1 proposes building more roads, Policy 2 proposes creating bus lanes on existing roads, and Policy 3 proposes license plate restrictions based on weekdays. (For the sake of the argument assume that the policies are mutually exclusive.) Within the population we observe that proponents of Policy 1 feel that Policy 3 limits the freedom of drivers, and proponents of Policy 3 feel that Policy 1 encourages car use instead of public transport, and the two groups feel they have no common ground. Therefore Policy 1 and Policy 3 are incompatible (in the sense used in this paper), and Policy 2 is compatible with both of the others.

In such scenarios, compatibility of opinions is subjective to some degree, and should be decided on a case-by-case basis, after observing the group's internal dynamics. In another city, it may be that Policy 2 is incompatible with others, due to the past history (feuds, etc.) of the group.

\subsection{Comparison with other models}
\label{subsec:model-comparison}
In this section the model described above is compared to existing similar models. Most of these models have been extended in different levels of sophistication, e.g. using heterogeneous agents or complicated network topologies. Here we consider only their fundamental features.

In Axelrod's model for the dissemination of culture~\cite{Axelrod1997}, agents' opinions are vectors composed of integers. At each step one agent and one of its neighbors are selected randomly. The two interact with a probability proportional to the number of opinions they agree on. If interaction takes place, the active agent copies one of the differing opinions of the neighbor.

The only common feature between the Axelrod model and the current model is that both are non-compromising. One agent adopts the opinion of the other as it is, instead of both of them converging to each other. The Axelrod model does not use bounded confidence; instead, the proximity of opinions is measured by the number of identical opinions. Also, the Axelrod model makes sense only with vector opinions, which are out of the scope of this study.

The model by Laguna et al.~\cite{LagunaEtAl2003} uses vector opinions with binary values in each component. It adds bounded confidence to Axelrod's model by making two agents interact only if the Hamming distance between them (the number of different components) is less than a threshold.

The classical Voter Model~\cite{CasForLor2010} is a very simplified version of the Axelrod model. It assigns a scalar binary opinion to each agent. At every step, one agent is chosen randomly and it adopts the opinion of a random neighbor. As there are only two opinions, bounded confidence does not apply and any agent can interact with any other. The Constrained Voter Model~\cite{VazquezRedner2004} is an extension that adds an intermediate opinion. Agents can be leftists, centrists or rightists. Leftists and rightists do not interact with each other, but only with centrists. These models are special cases of the model studied here: The classical Voter Model is obtained by setting $Q=2$, the Constrained Voter Model is obtained by setting $Q=3$, $d=1$.

Mobilia~\cite{Mobilia2011} introduces a control parameter $\alpha$ to the Constrained Voter model so that when $\alpha>0$ the absorbing state is more likely to be composed of extremists, and when $\alpha<0$ it is more likely to be centrist. This can be implemented in the current model by making the probability of conversion $p$ a function of interacting opinions.

Although there is not a simple equivalence between the Deffuant-Weisbuch model~\cite{DeffuantEtAl2000,WeisbuchEtAl2002} and the present model, some similarities exist. In both models consensus is more likely as the tolerance parameter increases. Simulations~\cite{DeffuantEtAl2000,Fortunato2004} indicate that for continuous opinions, the population almost always ends up in consensus when the tolerance parameter is larger than 0.5, or in discrete terms, larger than $Q/2$. In this paper's model, consensus is the only possible outcome if and only if the tolerance parameter $d$ is $Q-1$, its largest possible value. 

Discretized versions of the Deffuant-Weisbuch model round the opinions to the nearest integer (e.g.~\cite{Stauffer2005}). In case of binary opinions, one agent adopts the opinion of the other with probability equal to the convergence parameter of the model~\cite{DeffuantEtAl2000,StaufferEtAl2004}. With multiple opinions, if agents' opinions differ only by one, one agent can adopt the opinion of the other agent with probability 0.5~\cite{Assmann2004}.

Agents in the Continuous Opinions - Discrete Actions (CODA) Model~\cite{Martins2008} carry only binary (or ternary~\cite{Martins2009}) values of opinions, and an agent converts to the rival opinion with a probability that is updated at each time step. Effectively, agents have a memory of past interactions. The CODA model is similar to Urbig's model~\cite{Urbig2003} in its observance of verbalization limits. In this paper's model currently there is not an individual conversion probability associated with each agent, so it is not as general as the CODA model.

\section{Dynamical features}
\label{sec:dynamics}

\begin{figure}
\includegraphics{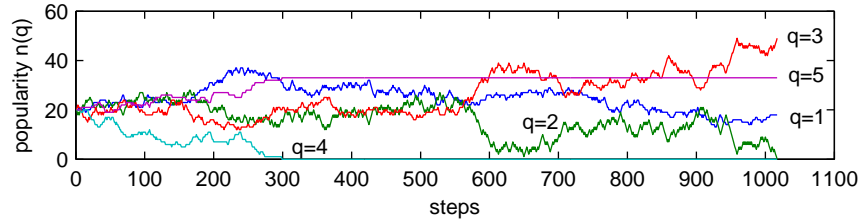}
\caption{\label{fig:pop_vs_time}One realization of the popularities of five opinions versus time. 100 agents, confidence bound 1. Fully mixed system, unbiased interactions.}
\end{figure}

We define the \emph{popularity} $n_q$ of opinion $q$ as the number of agents with opinion $q$. Figure~\ref{fig:pop_vs_time} shows a typical time series of popularities of five opinions. The opinions are initially distributed uniformly over 100 agents. The confidence bound $d$ is equal to one, and there is no bias in the interactions (both agents in a chosen pair are equally likely to be converted). In this particular simulation we see that opinion~4 loses all of its followers by step 300. After that time agents with opinion~5 cannot interact with any other agent, so the number of agents with opinion~5 does not change any more. Similarly, agents with opinions 1, 2 and 3 keep interacting until the number of followers of opinion~2 drops to zero at step 1017. After that, no more interactions are possible as the surviving opinions 1, 3 and 5 differ by two, which is more than the confidence bound $d=1$. 

\subsection{Agents' motion}
\label{sub:dynamics-motion}

The agents move randomly in the opinion space as they interact with each other. Even if an agent starts with an extreme opinion, in time it can move to farther, initially incompatible opinions, as long as there are other compatible agents with whom it can interact and use as a stepping stone.

An opinion that loses all of its followers is forever lost (unless the model involves some random noise that repopulates it). Absence of an opinion may block the migration of agents in the opinion space. For example, if there are no agents with opinion 3 and if the tolerance parameter $d$ is equal to one, there will be no crossover between opinions 2 and 4 because such an interchange would require interaction with opinion 3 first. If $d=2$ such a crossover is possible as agents with opinions 2 and 4 can directly interact.

\subsection{Relation to the Gambler's Ruin problem}
\label{sub:dynamics-gamblerruin}

With two opinions, the dynamics is equivalent to the classical \textit{Gambler's Ruin} problem~\cite{Feller1968}, where each player gains or loses one unit with equal probability. Here, players are opinions that gain or lose followers. Even if one starts with several opinions, toward the end of the run, two opinions may be left within each other's confidence bound, reducing the system to the classical problem.

The multiple player Gambler's Ruin is described by Ross~\cite{Ross2009} as follows: $Q$ players (opinions) begin with different amounts of capital (popularity) $n_q$ ($q=1\ldots Q$) at their disposal. At each step two players are chosen to play one round and each player is equally likely to win one unit. Any player whose capital drops to zero is eliminated. The run continues until all units are accumulated by one player. So, this game is equivalent to this paper's model with unbiased interactions and $d=Q-1$.

Under these conditions, regardless of how pairs of players (opinions) are chosen, Ross proves that:
\begin{enumerate}
 \item The expected number of interactions is finite, and given by $\left(N^2 - \sum_{q=1}^Q n_q^2\right)/2$.
 \item The probability that opinion $q$ is the ultimate winner (consensus opinion) is $n_q/N$.
 \item The expectation value of the number of interactions involving opinion $q$ is $n_q(N-n_q)$.
 \item The expectation value of the number of interactions involving only opinions $q$ and $q^\prime$ is $n_q n_{q^\prime}$.
\end{enumerate}

The related ``$N$-Tower problem'' considers the game ended when only \emph{one} of the players loses all of its capital. Bruss et al.~\cite{BrussEtAl2003} give the probability distribution, mean and variance of the absorbing time for the case of three towers (opinions).

\subsection{Absorbing states}
\label{sub:dynamics-absorbing}
After repeated interactions the system eventually reaches an absorbing state where no further change is possible. This can happen in two ways:
\begin{enumerate}
 \item No agent is compatible with any of its neighbors. This case happens when the network of agents has low connectivity (a few connections per agent). It is discussed in more detail in Section~\ref{sec:network}.
 \item An opinion $q$ has non-zero popularity and all compatible opinions $q-d,\ldots, q+d$ have zero popularity, so no interactions are possible. When the agents' network is completely connected, this is the necessary and sufficient condition for absorbing state.
\end{enumerate}

If there is a bias in the interaction that favors less-popular opinions, or if there is random noise that repopulates empty populations, the system may not be able to arrive at an absorbing state.

We call an opinion \textit{stationary} if all opinions compatible with it have zero popularity. In the unbiased case and with some biases (see subsection~\ref{subsec:fully-biased}) the system almost surely ends up in an absorbing state composed of several stationary opinions. Each stationary opinion is separated by at least $d+1$ empty opinions. Because of the stochastic nature of the dynamics, the same initial conditions may lead to a different set of end states in each simulation, so neither the stationary opinions themselves nor their number can be predicted in advance. However, we can determine the maximum number $m$ of stationary opinions by considering the tightest possible arrangement of absorbing states. If $Q$ is the total number of opinions and $d$ is the confidence bound, starting from opinion one, the tightest stationary arrangement is obtained with steps of $d+1$:
\begin{equation}
1,\ 1+(d+1),\ 1+2(d+1),\ \cdots\ 1+(m - 1)(d+1) \leq Q.
\end{equation}
Alternatively, one can start at opinion $Q$ and make another tight arrangement by going backwards with steps of $d+1$. Either case has $m$ states, where $m$ is the maximum number of stationary opinions, given by 
\begin{equation}
\label{eq:maxstationary}
m=\left\lfloor\frac{Q-1}{d+1}\right\rfloor + 1,
\end{equation} 
where $\lfloor\cdot\rfloor$ denotes the integer part. In particular if $d=Q-1$, only one stationary opinion (consensus) may exist. For $d<Q-1$, there may be one or more stationary opinions.

\subsection{Biased interactions}
\label{subsec:fully-biased}

In order to make the model more realistic we can introduce a bias in the interaction. This can be achieved by making the conversion probability $p$ as a function of agent properties.

When agents are identical, $p$ can depend on interacting opinions $q$, $q^\prime$. With proper choice of parameters, such a bias can reproduce Mobilia's~\cite{Mobilia2011} version of the constrained voter model. 

Alternatively, the bias could depend on the popularities $n$, $n^\prime$ of interacting opinions $q$, $q^\prime$, respectively. For example, if an opinion is shared by more people, it could gain new followers more easily. This realization of the Matthew effect can be modelled in many ways, including:
\begin{enumerate}
 \item Pairwise majority bias: If $n>n^\prime$, $q$ is favored with probability
\begin{equation}
\label{eq:bias_maj_pair}
p= 0.5 + 0.5|n-n^\prime|/(n+n^\prime).
\end{equation}
\item Global majority bias: If $n>n^\prime$, $q$ is favored with probability
\begin{equation}
\label{eq:bias_maj_global}
p= 0.5 + 0.5|n-n^\prime|/N
\end{equation}
where $N$ is the total number of agents.
\end{enumerate}
The pairwise majority bias considers only the relative difference between the two interacting opinions. The global majority bias puts this difference in a system-wide perspective. With the latter, the unfavored agent is not much impressed if the other agent's opinion is not very popular over the entire population. The difference between these two biases is significant only when there are many competing opinions. Any bias to majority will accelerate the convergence to an absorbing state.

Strictly speaking, as the popularity is a global variable, its use is not compatible with zero-intelligence agents. Still, in a completely connected society where everybody knows everybody, it is not unreasonable that every agent has an idea about the popularities of opinions.

\begin{figure}
\includegraphics[scale=0.8]{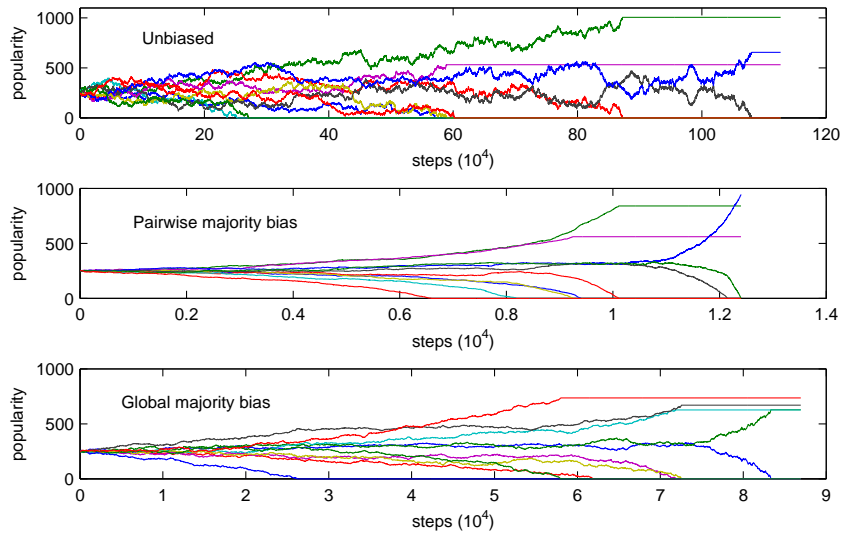}
\caption{\label{fig:biascomparison}A comparison of time evolution of popularities under no bias, pairwise majority bias and global majority bias. 5000 agents, 20 opinions, confidence bound 1. Only some opinions are shown.}
\end{figure}

Figure~\ref{fig:biascomparison} compares typical time series under different bias conditions. Under pairwise majority bias, the time series is much smoother and reaches the end state rapidly.

One can also set up a bias to minority opinions. For example, if the popularity of an opinion is too high, agents may be set up to ``seek novelty'' so that interactions are biased toward less popular opinions. In such models stationary opinions may not exist, but a dynamical equilibrium may be possible. Such biases are not studied in this paper.

\section{Fully mixed system}
\label{sec:fully}
In a ``fully mixed'' (homogeneous) system, all agents have the same features, and pairwise interactions between agents do not depend on individual properties. Furthermore, any agent can interact with all other agents as long as the bounded confidence condition is satisfied. A fully mixed system is a model for a small community where individuals are free to move around and discuss.

\subsection{Markov chain analysis}
\label{subsec:markov}
Define $s=(n_1, n_2, \ldots,n_Q)$ to be a state of the system. Let $X_t$ be the state of the system at step $t$, and let $\mathrm{Pr}\{X_t=(n_1, n_2, \ldots,n_Q)\}$ be the probability that at time $t$ opinion 1 has $n_1$ followers, opinion 2 has $n_2$ followers, etc. By the law of total probability we have:
\begin{equation}
 \mathrm{Pr}\{X_{t+1}=s\} = \sum_{s^\prime}\mathrm{Pr}\{X_{t+1}=s|X_t=s^\prime\} \mathrm{Pr}\{X_t=s^\prime\}
\end{equation}
where the sum is over all states $s^\prime=(n_1^\prime, n_2^\prime, \ldots,n_Q^\prime)$.

At each time step a pair of agents with compatible opinions $q$ and $q^\prime$ ($q \neq q^\prime$) are selected. There are only two possible transitions: $\lbrace n_q\rightarrow n_q-1,\ n_{q^\prime}\rightarrow n_{q^\prime}+1\rbrace$, or, $\lbrace n_q\rightarrow n_q+1,\ n_{q^\prime}\rightarrow n_{q^\prime}-1 \rbrace$. Let $s$ be the state that the popularity of opinion $i$ is $n_i$. Let $s^\prime$ be the state where the popularity of $q$ is $n_q-1$, the popularity of $q^\prime$ is $n_{q^\prime}+1$, and the remaining popularities are the same as in $s$. For both of these transitions, the transition rates are given by:
\begin{equation}
\mathrm{Pr}\{X_{t+1}=s^\prime|X_t=s\} = p\frac{n_q n_{q^\prime}}{M(\{n_i\})}
\end{equation}
where $p=0.5$ for the unbiased interaction, and
\begin{equation}
 M(\{n_i\}) = \sum_{k=1}^{Q-1} \sum_{k^\prime = k+1}^{\min(k+d,Q)} n_k n_{k^\prime}
\end{equation}
is the number of all compatible pairs of agents in a given configuration ${n_i}$. Note that $M(\{n_i\})=0$ for absorbing states.

Then the stochastic dynamics of opinion popularities is described by the following coupled linear system:
\begin{eqnarray}
\label{eq:prob_general}
\mathrm{Pr}\{X_{t+1}=s\} &=& \nonumber\\
&& \sum_{q=1}^Q\sum_{q^\prime =\max(1,q-d)}^{\min(q+d,Q)} p \frac{(n_q - 1)(n_{q^\prime} +1)} {M(s^\prime)} \Theta(n_q -1) \mathrm{Pr}\{X_t=s^\prime\} \nonumber\\
&+& \delta_{M(s),0} \mathrm{Pr}\{X_t=s\}
\end{eqnarray}
The sum is over all ordered compatible opinion pairs $(q,q^\prime)$ where $q^\prime\neq q$. The step function $\Theta(n_q-1)$ ensures that opinion $q$ has at least one follower. The last term sets the absorbing state probabilities using the Kronecker delta symbol.

With $N$ agents and $Q$ opinions, the number of possible states is
\begin{equation}
 \frac{(N+Q-1)!}{(Q-1)!N!}
\end{equation}
which is also the number of equations. With a proper enumeration of states, this linear system can be converted to a matrix-vector equation. The matrix is a Markov matrix, so all of its eigenvalues are less than or equal to 1. The eigenvectors corresponding to unit eigenvalues are absorbing states.

The eigensystem of the Markov matrix can be determined numerically, yielding the relative probability of each absorbing state, as well as transient properties like absorption times. However, as the matrix size grows as $O(N^{Q-1})$, this approach is feasible for small systems only. A direct agent-based simulation is needed for larger systems.

The probabilities are expressed so that there is exactly one conversion between opinions, in accordance with the agent-based simulations. It is possible to modify the model so that at each time step any two agents are chosen randomly, compatible or not. In that case the system will have the same absorbing states, but its convergence will be much slower. For this modification we need to set 
$M(s^\prime)= N(N-1)$
 in Eq.(\ref{eq:prob_general}), and replace the last term of the same equation with
\begin{equation}
\sum_{(q,q^\prime)}\frac{n_q n_{q^\prime}}{N(N-1)}\mathrm{Pr}\{X_t = s\}
\end{equation}
where the sum is over all ordered opinion pairs $(q,q^\prime)$ that do not result in a change of state. For example, if $Q=3$ and $d=1$, the sum will be evaluated with pairs (1,3), (3,1), (1,1), (2,2), and (3,3).
\subsection{Three opinions}
\label{subsec:three_ops}
Here we apply Eq.(\ref{eq:prob_general}) to the case of three opinions in a population of $N$ agents with unbiased interactions.

Let the confidence bound $d$ be 1. By inspection it is seen that there are $N+2$ absorbing states: $(0, N, 0)$, and $(i, 0, N-i)$ where $i=0\ldots N$. Compatible opinion pairs are (1,2), (2,1), (2,3) and (3,2), and $M(x,y,z)=xy+yz$. Then, Eq.(\ref{eq:prob_general}) takes the form:
\begin{eqnarray}
 \mathrm{Pr}\{X_{t+1}=(n_1,n_2,n_3)\} & = & \frac{1}{2} \frac{(n_1-1)(n_2+1) \Theta (n_1-1)}{(n_1-1)(n_2+1)+(n_2+1)n_3} \mathrm{Pr}\{X_t=(n_1-1, n_2+1, n_3)\} \nonumber\\
 & + & \frac{1}{2} \frac{(n_1+1)(n_2-1) \Theta (n_2-1)}{(n_1+1)(n_2-1)+(n_2-1)n_3} \mathrm{Pr}\{X_t=(n_1+1, n_2-1, n_3)\} \nonumber\\
 & + & \frac{1}{2} \frac{(n_2-1)(n_3+1) \Theta (n_2-1)}{n_1(n_2-1)+(n_2-1)(n_3+1)} \mathrm{Pr}\{X_t=(n_1, n_2-1, n_3+1)\} \nonumber\\
 & + & \frac{1}{2} \frac{(n_2+1)(n_3-1) \Theta (n_3-1)}{n_1(n_2+1)+(n_2+1)(n_3-1)} \mathrm{Pr}\{X_t=(n_1, n_2+1, n_3-1)\} \nonumber\\
 & + & \delta_{M(n_1,n_2,n_3),0} \mathrm{Pr}\{X_t = (n_1, n_2, n_3)\}
\end{eqnarray}
As $n_3 = N-n_1-n_2$, a state can be specified with $n_1$ and $n_2$ alone. We list the states such that $n_2$ varies from 0 to $N-n_1$, and $n_1$ varies from 0 to $N$, and we define $i(n_1, n_2, n_3)$ to be the position of state $(n_1,n_2,n_3)$ on this list ($n_3$ is kept for notational convenience). The list goes as follows:
\begin{eqnarray}
i(0,0,N) &=& 1 \nonumber\\
i(0,1,N-1) &=& 2\nonumber\\
&\hdots& \nonumber\\
i(0,N,0) &=& N+1 \nonumber\\
i(1,0,N-1) &=& N+2 \nonumber\\
&\hdots& \nonumber\\
i(N,0,0) &=& (N+2)(N+3)/2
\end{eqnarray}

Then, the index function must have the following form:
\begin{equation}
 i(n_1, n_2, n_3) = (N-2)n_1 - \frac{1}{2}n_1(n_1+1) + n_2 + 1
\end{equation}
Now we define a column vector $\mathbf{p}^t = [p_i^t]$ such that $p^t_{i(n_1,n_2,n_3)} = \mathrm{Pr}\{X_t=(n_1,n_2,n_3)\}$. Then the Markov matrix $A=[a_{ij}]$ is given by
\begin{eqnarray}
 a_{i(n_1,n_2,n_3),j} & = & \frac{1}{2} \frac{(n_1-1)(n_2+1)\Theta (n_1-1)}{(n_1-1)(n_2+1)+(n_2+1)n_3}  \delta_{i(n_1-1,n_2+1,n_3), j} \nonumber\\
 & + & \frac{1}{2} \frac{(n_1+1)(n_2-1)\Theta (n_2-1)}{(n_1+1)(n_2-1)+(n_2-1)n_3} \delta_{i(n_1+1,n_2-1,n_3), j} \nonumber\\
 & + & \frac{1}{2} \frac{(n_2-1)(n_3+1) \Theta (n_2-1)}{n_1(n_2-1)+(n_2-1)(n_3+1)} \delta_{i(n_1,n_2-1,n_3+1), j} \nonumber\\
 & + & \frac{1}{2} \frac{(n_2+1)(n_3-1) \Theta (n_3-1)}{n_1(n_2+1)+(n_2+1)(n_3-1)} \delta_{i(n_1,n_2+1,n_3-1), j} \nonumber\\
 & + & \delta_{i(n_1,n_2,n_3), j}\delta_{M(n_1,n_2,n_3), 0}
\end{eqnarray}
so that $\mathbf{p}^{t}=A\mathbf{p}^{t-1}$.

Let $\lambda_1, \lambda_2, \ldots$ be the eigenvalues of $A$, arranged in decreasing order, and let $\mathbf{v}_1, \mathbf{v}_2, \ldots$ be the associated eigenvectors. As the system has $N+2$ absorbing states, it holds that $\lambda_1 = \ldots = \lambda_{N+2} = 1$, and all other eigenvalues are smaller than unity. 

Matrix $A$ is diagonalizable, so the equation can be written as:
\begin{equation}
 \mathbf{p}^{t}=VDV^{-1}\mathbf{p}^{t-1} = VD^{t}V^{-1}\mathbf{p}^0
\end{equation}
where $V$ is the matrix whose columns are the eigenvectors of $A$, $D$ is the diagonal matrix whose entries are the eigenvalues of $A$, and $\mathbf{p}^0$ is the initial probability vector of states. The vector $V^{-1}\mathbf{p}^0\equiv\mathbf{c}$ gives the initial vector in the eigenvector basis. Then it holds that
\begin{eqnarray}
 \mathbf{p}^{t} &=& VD^{t}\mathbf{c} \nonumber \\
&=& \lambda_1^{t}c_1\mathbf{v}_1 + \lambda_2^{t}c_2\mathbf{v}_2 + \ldots 
\end{eqnarray}
where $c_i$ is the $i$-th component of $\mathbf{c}$. As $t\rightarrow\infty$, all terms with $\lambda_i<1$ will vanish, and the steady state probability vector will be a linear combination of absorbing state eigenvectors:
\begin{equation}
\label{eq:ss_prob}
 \mathbf{p}^\infty = c_1\mathbf{v}_1 + c_2\mathbf{v}_2 + \ldots + c_{N+2}\mathbf{v}_{N+2}
\end{equation}
Therefore, $c_i$ is the probability that the system ends up in the absorbing state corresponding to $\mathbf{v}_i$. Numerical results for these absorbing state probabilities are given section~\ref{subsec:absstate-prob}.

If the confidence bound $d$ is 2, compatible opinion pairs are (1,2), (1,3), (2,1), (2,3), (3,1) and (3,2), and $M(x,y,z) = xy + yz + zx$. Then the Markov matrix is given by:
\begin{eqnarray}
 a_{i(n_1,n_2,n_3),j}
 & = & \frac{1}{2} \frac{(n_1-1)(n_2+1)\Theta (n_1-1)}{(n_1-1)(n_2+1) + (n_2+1)n_3 + (n_1-1)n_3}  \delta_{i(n_1-1,n_2+1,n_3), j} \nonumber\\
 & + & \frac{1}{2} \frac{(n_1+1)(n_2-1)\Theta (n_2-1)}{(n_1+1)(n_2-1) + (n_2-1)n_3 + (n_1+1)n_3} \delta_{i(n_1+1,n_2-1,n_3), j} \nonumber\\
 & + & \frac{1}{2} \frac{(n_2-1)(n_3+1) \Theta (n_2-1)}{n_1(n_2-1) + (n_2-1)(n_3+1) + n_1(n_3+1)} \delta_{i(n_1,n_2-1,n_3+1), j} \nonumber\\
 & + & \frac{1}{2} \frac{(n_2+1)(n_3-1) \Theta (n_3-1)}{n_1(n_2+1) + (n_2+1)(n_3-1) + n_1(n_3-1)} \delta_{i(n_1,n_2+1,n_3-1), j} \nonumber\\
 & + & \frac{1}{2} \frac{(n_1+1)(n_3-1) \Theta (n_3-1)}{(n_1+1)n_2 + n_2(n_3-1) + (n_1+1)(n_3-1)} \delta_{i(n_1+1,n_2,n_3-1), j} \nonumber\\
 & + & \frac{1}{2} \frac{(n_1-1)(n_3+1) \Theta (n_1-1)}{(n_1-1)n_2 + n_2(n_3+1) + (n_1-1)(n_3+1)} \delta_{i(n_1-1,n_2,n_3+1), j} \nonumber\\
 & + & \delta_{i(n_1,n_2,n_3), j}\delta_{M(n_1,n_2,n_3), 0}
\end{eqnarray}
which can be analyzed as described above.

Using this method, one can set up explicit Markov matrices for arbitrary $Q$ and $d$. However, the explicit form of the index function $i(\{n_i\})$ that maps the linear index to states is too difficult to determine for general $Q$ and $N$. Instead of a functional form, one can use a lookup table for this map when setting up the matrix in software.

Matlab code that evaluates the absorbing state probabilities for $Q=3$ and $Q=4$ using Markov matrices is given in the supplement to this paper.

\subsection{Absorbing state probabilities: Consensus or polarization?}
\label{subsec:absstate-prob}
The system is said to be in \textit{consensus} if all agents finally end up in the same opinion, \textit{polarized} if there are only two remaining opinions (necessarily incompatible with each other), and \textit{fragmented} otherwise.

For $Q=2$, only consensus is possible. If $Q=3$ and $d=1$, consensus or polarization is possible. Fragmentation (three or more stationary opinions) is possible only for $Q\ge 5$.

In general, there can be at most $ \lfloor(Q-1)/(d+1)\rfloor + 1$ opinions in the absorbing state. If $d=Q-1$ there can be only consensus. If $(Q-3)/2 < d < Q-1$ holds, the system can reach consensus or polarization, but not fragmentation.

Using the Markov matrix, the probability of an absorbing state can be determined as described in Section~\ref{subsec:three_ops}. For $Q=3$ and $d=1$, possible absorbing states are $(0,N,0)$ and $(i,0,N-i)$, where $i=0\ldots N$. Using a uniform initial condition ($n_1=n_2=n_3=N/3$), the matrix is numerically analyzed for $N=15,30,60,$ and 99. In each case, it is found that the probability of the absorbing state $(0,N,0)$ is $1/3$, within numerical precision, independent of $N$.

\begin{figure}
\includegraphics[scale=0.75]{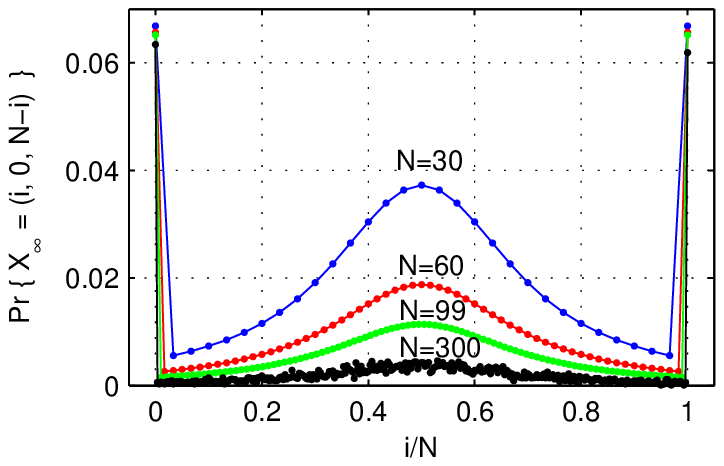}
\includegraphics[scale=0.75]{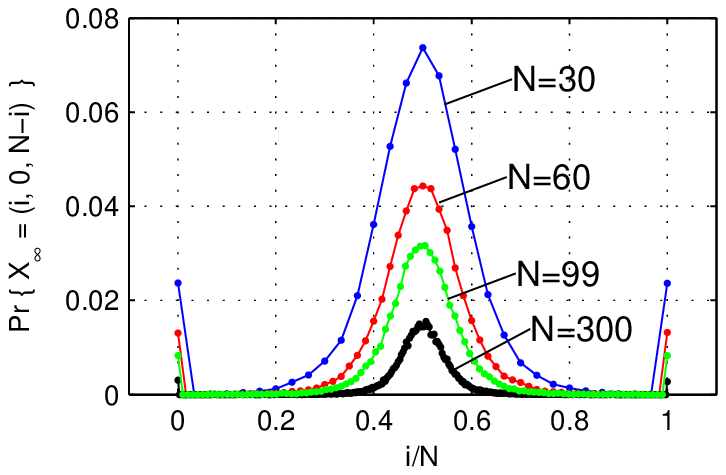}
\caption{\label{fig:Q3d1_abs_states}The probability the absorbing state $(i,0,N-i)$ for $Q=3$, $d=1$. Left: Without bias. The curve for $N=300$ is obtained by direct agent-based simulation, the rest by Markov analysis. Right: With bias to pairwise majority (Eq.~\ref{eq:bias_maj_pair}); all curves obtained by direct simulation.}
\end{figure}

Figure~\ref{fig:Q3d1_abs_states} shows the probability of the absorbing state $(i, 0, N-i)$ as a function of $i$, for various $N$, for unbiased and biased interactions. Bias is taken to be pairwise majority bias, as given in~(\ref{eq:bias_maj_pair}).

In the unbiased case (left panel in Fig.~\ref{fig:Q3d1_abs_states}) the consensus states $(0,0,N)$ and $(N,0,0)$ are more likely than polarized states. The probability of consensus states is about 0.065, and varying only slowly with $N$.

The case of $N=300$ could not be analyzed with the Markov approach as the matrix is too big. This curve is obtained by using direct agent-based simulation, averaging over $10^4$ runs. In the biased case all calculations are agent-based, averaging over $10^5$ runs.

In all cases the curves have a peak in the middle, indicating that an equal division $(N/2, 0, N/2)$ is the most likely polarization. In the unbiased interaction the curve becomes flatter as $N$ increases, suggesting that all polarization states are equally likely as $N\rightarrow\infty$. In the biased interaction, however, the central peak persists, therefore the near-equal polarizations are dominant.

\begin{table}
 \caption{\label{table:n_ops_prob}Probabilities of the number of opinions in the absorbing state with $N\approx 100$ agents, averaging over $10^5$ runs.}
\begin{tabular}[t]{|l|c|c|c|c|c|c|c|c|c|c|c|}
\cline{4-7} \cline{9-12} 
\multicolumn{1}{l}{} & \multicolumn{1}{c}{} &  & \multicolumn{4}{c|}{no bias} &  & \multicolumn{4}{c|}{with bias}\tabularnewline
\cline{1-2} \cline{4-7} \cline{9-12} 
$Q$ & $d$ &  & \textbf{1} & \textbf{2} & \textbf{3} & \textbf{4} &  & \textbf{1} & \textbf{2} & \textbf{3} & \textbf{4}\tabularnewline
\cline{1-2} \cline{4-7} \cline{9-12} 
3 & 1 &  & 0.4634 & 0.5366 & - & - &  & 0.4644 & 0.5356 & - & -\tabularnewline
\cline{1-2} \cline{4-7} \cline{9-12} 
4 & 1 &  & 0.1255 & 0.8745 & - & - &  & 0.0079 & 0.9921 & - & -\tabularnewline
\cline{2-2} \cline{4-7} \cline{9-12} 
 & 2 &  & 0.6735 & 0.3265 & - & - &  & 0.6142 & 0.3858 & - & -\tabularnewline
\cline{1-2} \cline{4-7} \cline{9-12} 
5 & 1 &  & 0.0236 & 0.6770 & 0.2995 & - &  & 0.0001 & 0.6294 & 0.3704 & -\tabularnewline
\cline{2-2} \cline{4-7} \cline{9-12} 
 & 2 &  & 0.4042 & 0.5958 & - & - &  & 0.3134 & 0.6866 & - & -\tabularnewline
\cline{2-2} \cline{4-7} \cline{9-12} 
 & 3 &  & 0.7831 & 0.2169 & - & - &  & 0.6967 & 0.3033 & - & -\tabularnewline
\cline{1-2} \cline{4-7} \cline{9-12} 
6 & 1 &  & 0.0029 & 0.3309 & 0.6662 & - &  & $<10^{-5}$ & 0.1854 & 0.8146 & -\tabularnewline
\cline{2-2} \cline{4-7} \cline{9-12} 
 & 2 &  & 0.1897 & 0.8103 & - & - &  & 0.0216 & 0.9784 & - & -\tabularnewline
\cline{2-2} \cline{4-7} \cline{9-12} 
 & 3 &  & 0.5680 & 0.4320 & - & - &  & 0.4649 & 0.5351 & - & -\tabularnewline
\cline{2-2} \cline{4-7} \cline{9-12} 
 & 4 &  & 0.8425 & 0.1575 & - & - &  & 0.7529 & 0.2471 & - & -\tabularnewline
\cline{1-2} \cline{4-7} \cline{9-12} 
7 & 1 &  & 0.0002 & 0.1149 & 0.7151 & 0.1698 &  & $<10^{-5}$ & 0.0116 & 0.7474 & 0.2410\tabularnewline
\cline{2-2} \cline{4-7} \cline{9-12} 
 & 2 &  & 0.0839 & 0.7878 & 0.1283 & - &  & 0.0032 & 0.7714 & 0.2254 & -\tabularnewline
\cline{2-2} \cline{4-7} \cline{9-12} 
 & 3 &  & 0.3832 & 0.6168 & - & - &  & 0.2532 & 0.7468 & - & -\tabularnewline
\cline{2-2} \cline{4-7} \cline{9-12} 
 & 4 &  & 0.6731 & 0.3269 & - & - &  & 0.5615 & 0.4385 & - & -\tabularnewline
\cline{2-2} \cline{4-7} \cline{9-12} 
 & 5 &  & 0.8842 & 0.1158 & - & - &  & 0.7913 & 0.2087 & - & -\tabularnewline
\cline{1-2} \cline{4-7} \cline{9-12} 
\end{tabular}
\end{table}

Table~\ref{table:n_ops_prob} lists the probabilities of finding one opinion (consensus), two opinions (polarization), and three and four opinions (fragmentation) in the absorbing state, for $Q$ up to 7. The probabilities are found by agent-based simulation, averaging over $10^5$ runs. The system is initialized with equal popularities for all opinions. Biased simulations use pairwise majority bias. The number of agents is either 100, or the largest number smaller than 100 divisible by $Q$. For any $Q$, runs with $d=Q-1$ always result with consensus, so they are omitted in the table.

The results are not sensitive to the number of agents. Larger populations (up to 1000) change the values only by about 0.01, usually less.

The constrained voter model as described by Vazquez and Redner~\cite{VazquezRedner2004} is equivalent to the case $Q=3$, $d=1$, unbiased interactions. The authors give the probabilities of polarized states and consensus states as functions of initial densities. Applying these results to the uniform initial distribution, one finds that the probability of a polarized state is 0.5377, probability of consensus on either extreme is 0.0645, and the probability of consensus on the middle opinion is $1/3$. These values agree with the results given above and in Table~\ref{table:n_ops_prob}.

The chance of reaching a consensus varies greatly with the confidence bound $d$. When the confidence bound is small, the population is more likely to end up in polarization, or even fragmentation, instead of reaching a consensus. This is a common feature of bounded-confidence models: It is harder to reach a consensus when individuals do not have a high tolerance for different opinions. Narrow-mindedness splits society into several factions that do not speak to each other.

Still, the population is usually not as much fragmented as it can possibly be. The entries for $Q=5$ and $Q=7$ with $d=1$ show that the probability of the most fragmented state is smaller than a less-fragmented one.

When interactions involve bias to pairwise majority, the same properties hold, but the system tends to be slightly more polarized or fragmented. Again, these results are not sensitive to the number of agents.

\subsection{Absorption time distribution}
\label{subsec:fully-eqbtime}
The absorption time (number of steps until there are no more opinion interchanges) is of particular interest. With unbiased interactions, the expectation value of absorption time can be determined using the results of Ross~\cite{Ross2009} as described in section~\ref{sub:dynamics-gamblerruin}.

Let $n_i,n_j$ be the initial popularities of opinions $i$ and $j$, respectively. The expected number of interactions between these opinions is $n_i n_j$. Then the expected absorption time is given by the sum:
\begin{equation}
 \mathrm{E}[t_{\rm abs}] = \sum_{i,j} n_i n_j,
\end{equation}
where the sum is over all compatible pairs of opinions such that $i<j$ and $1\leq |i-j| \leq d$.

When the initial distribution is uniform such that $n_i = N/Q$ for all $i$, this expression reduces to:
\begin{equation}
 \mathrm{E}[t_{\rm abs}] = \frac{d(2Q-d-1)}{2Q^2}N^2.
\end{equation}
Here, $d(2Q-d-1)/2$ is the number of compatible opinion pairs for given $Q$ and $d$. This number [call it $c(Q,d)$] can be found recursively: It holds that $c(Q,d) = c(Q,d-1) + Q-d$, because to go from tolerance $d-1$ to tolerance $d$ we need to connect opinion pairs $(1,d),(2,d+1),\ldots,(Q-d,Q)$, adding $Q-d$ pairs to the sum. Using the initial condition $c(Q,1)=Q-1$ we get the result above.

The expected absorption time scales as $N^2$, decreases with $Q$ for fixed $d$, and increases monotonically with $d$. Narrow-minded agents (small $d$) require fewer steps to reach an absorbing state because they have fewer people to talk to.

\begin{figure}
\includegraphics[scale=0.75]{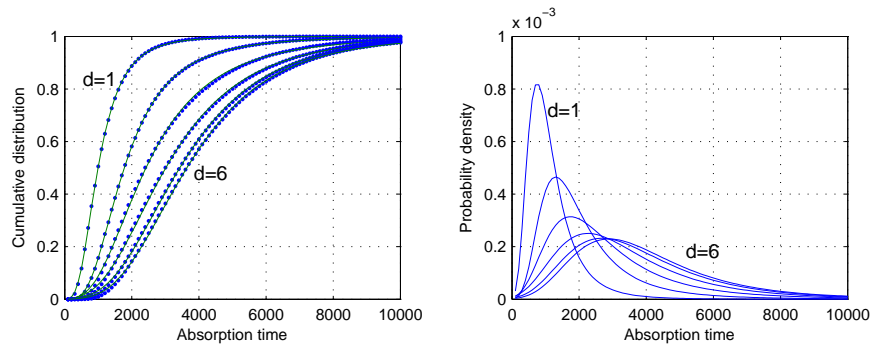}
\caption{\label{fig:timedist}Left: Cumulative distribution of absorption time for $N=98$, $Q=7$ and $d=1\ldots 6$, under unbiased interaction. Points are simulation results and curves are nonlinear fits to the Generalized Extreme Value distribution. For all fits $R^2>0.9996$.  Right: Probability density for absorption time (derivative of the fitted curves on the left).}
\end{figure}

The absorption time distribution gives more information than the expectation value. Figure~\ref{fig:timedist} shows the distribution with $N=98$ agents and $Q=7$ opinions. The values are extracted from the same set of agent-based simulations used to produce Table~\ref{table:n_ops_prob}.

With such wide distributions, it is more practical to use the Cumulative Distribution Function (CDF) than to form a histogram (i.e., the probability density). Unlike a histogram, a CDF is always continuous, monotonically increasing, and not very sensitive to the number of runs used for averaging. Once a functional fit to the CDF is found, its derivative yields the probability density.

The distribution of absorption times is very well approximated by the Generalized Extreme Value (GEV) distribution \cite{Sornette2006,Beirlant2004}. The GEV distribution arises from the extreme value theory, a branch of statistics that deals with the maximum (or minimum) value from a set of i.i.d. random numbers. The theory is applied to model a wide variety of phenomena, including flood levels, rainfall, insurance claims, seismic events, and human life span. Since the absorbing state occurs for extremal popularity values (zero or maximum), it is reasonable that the GEV distribution appears in this problem, too.

The GEV distribution has the cumulative distribution function:
\begin{equation}
\label{eq:GEVF}
 F(t; m, a, \xi) = \exp \left[-\left(1+\xi\frac{t-m}{a}\right)_+^{-1/\xi}\right]
\end{equation}
where the notation $x_+$ indicates $\max (x,0)$. The distribution has a location parameter $m$, a scale parameter $a>0$ and a shape parameter $\xi$.

A minimum number of steps are required before any opinion is depleted of its followers; before that time the absorption probability is zero. Due to this lower limit on time, the GEV distribution slightly overestimates the simulation results at small time values ($t<1500$). When biased interactions are used, the fit to GEV is almost perfect even at small time values (not shown in figure).

\begin{table}
\caption{\label{table:GEV}Parameters for the GEV distribution determined with nonlinear fit. For most of the fits $R^2>0.999$. Bias is to pairwise majority.}
\begin{tabular}[t]{|l|c|c|c|c|c|c|c|c|c|}
\cline{4-6} \cline{8-10} 
\multicolumn{1}{l}{} & \multicolumn{1}{c}{} &  & \multicolumn{3}{c|}{no bias} &  & \multicolumn{3}{c|}{with bias}\tabularnewline
\cline{1-2} \cline{4-6} \cline{8-10} 
$Q$ & $d$ &  & $a$ & \textbf{$m$} & \textbf{$\xi$} &  & $a$ & \textbf{$m$} & \textbf{$\xi$}\tabularnewline
\cline{1-2} \cline{4-6} \cline{8-10} 
3 & 1 &  & 1127 & 1176 & 0.2674 &  & 46.03 & 110.5 & -0.0096\tabularnewline
\cline{2-2} \cline{4-6} \cline{8-10} 
 & 2 &  & 1441 & 2193 & 0.1618 &  & 41.90 & 164.6 & 0.0735\tabularnewline
\cline{1-2} \cline{4-6} \cline{8-10} 
7 & 1 &  & 454.1 & 822.3 & 0.1753 &  & 30.66 & 118.2 & -0.0282\tabularnewline
\cline{2-2} \cline{4-6} \cline{8-10} 
 & 2 &  & 811.1 & 1466 & 0.2287 &  & 37.56 & 159.6 & 0.0117\tabularnewline
\cline{2-2} \cline{4-6} \cline{8-10} 
 & 3 &  & 1197 & 1966 & 0.2156 &  & 41.65 & 179.6 & 0.0459\tabularnewline
\cline{2-2} \cline{4-6} \cline{8-10} 
 & 4 &  & 1483 & 2455 & 0.1407 &  & 46.32 & 195.3 & 0.0276\tabularnewline
\cline{2-2} \cline{4-6} \cline{8-10} 
 & 5 &  & 1593 & 2834 & 0.1035 &  & 48.00 & 207.8 & 0.0156\tabularnewline
\cline{2-2} \cline{4-6} \cline{8-10} 
 & 6 &  & 1610 & 3040 & 0.0960 &  & 46.38 & 215.8 & 0.0376\tabularnewline
\cline{1-2} \cline{4-6} \cline{8-10} 
\end{tabular}
\end {table}

Table~\ref{table:GEV} shows the parameters found by fitting the CDF of absorption time to the GEV distribution. The $R^2$ parameter is used to verify the goodness of fit; $R^2=1$ indicates a perfect fit. The curve fits the simulation very well: The minimum value of $R^2$ is 0.9983 with $Q=3$, $d=1$, unbiased; for all other cases $R^2>0.999$.

\subsection{Survival probability of opinions}
\label{subsec:fully-survival}
Given a specific opinion $q$, one may ask how likely it is that this opinion will survive until the absorbing state. Table~\ref{table:survivalprob} shows the survival probability of opinions for $Q=3$, 5, and 7, under unbiased interactions. The values are again obtained from the simulations used to produce Tables~\ref{table:n_ops_prob} and~\ref{table:GEV}. Initial opinion distribution is uniform.

\begin{table}
\caption{\label{table:survivalprob}Probability that a specific opinion survives in the absorbing state. Unbiased interactions.}
\begin{tabular}[t]{|l|c|c|c|c|c|c|c|c|c|}
\cline{1-2} \cline{4-10} 
$Q$ & $d$ &  & \textbf{1} & \textbf{2} & \textbf{3} & \textbf{4} & \textbf{5} & \textbf{6} & \textbf{7}\tabularnewline
\cline{1-2} \cline{4-10} 
3 & 1 &  & 0.6021 & 0.3322 & 0.6023 & - & - & - & -\tabularnewline
\cline{2-2} \cline{4-10} 
 & 2 &  & 0.3338 & 0.3342 & 0.3320 & - & - & - & -\tabularnewline
\cline{1-2} \cline{4-10} 
5 & 1 &  & 0.5787 & 0.3553 & 0.4075 & 0.3570 & 0.5773 & - & -\tabularnewline
\cline{2-2} \cline{4-10} 
 & 2 &  & 0.4184 & 0.2796 & 0.2004 & 0.2788 & 0.4187 & - & -\tabularnewline
\cline{2-2} \cline{4-10} 
 & 3 &  & 0.3057 & 0.2009 & 0.2027 & 0.1995 & 0.3081 & - & -\tabularnewline
\cline{2-2} \cline{4-10} 
 & 4 &  & 0.2003 & 0.2007 & 0.2001 & 0.1994 & 0.1995 & - & -\tabularnewline
\cline{1-2} \cline{4-10} 
7 & 1 &  & 0.5778  &  0.3562 &  0.3972 & 0.3919 & 0.3971 &  0.3569 & 0.5774\tabularnewline
\cline{2-2} \cline{4-10} 
 & 2 &  & 0.4087  & 0.2768 & 0.2152 & 0.2422 & 0.2176 & 0.2766 & 0.4072\tabularnewline
\cline{2-2} \cline{4-10} 
 & 3 &  & 0.3218 & 0.2354 & 0.1784 & 0.1429 & 0.1818 & 0.2352 & 0.3214\tabularnewline
\cline{2-2} \cline{4-10} 
 & 4 &  & 0.2593 & 0.1894 & 0.1425 & 0.1435 & 0.1435 & 0.1892 & 0.2595\tabularnewline
\cline{2-2} \cline{4-10} 
 & 5 &  & 0.2010 & 0.1421 & 0.1410 & 0.1430 & 0.1411 & 0.1432 & 0.2015\tabularnewline
\cline{2-2} \cline{4-10} 
 & 6 &  & 0.1426 & 0.1415 & 0.1435 & 0.1433 & 0.1424 & 0.1434 & 0.1433\tabularnewline
\cline{1-2} \cline{4-10} 
\end{tabular}
\end{table}

Probabilities for a given $Q$ and $d$ do not add up to 1 because more than one opinion can be present in the absorbing state. For the cases where $d=Q-1$ it can be predicted that the survival probability is $1/Q$ for each opinion~\cite{Ross2009} (see section~\ref{sub:dynamics-gamblerruin}). The values on the table are consistent with this prediction. 

The table shows that extreme opinions are more likely to be present in the absorbing state than central opinions. This should not be interpreted as agents being more likely to end up in extreme opinions. Actually, with unbiased interactions, popularities of stationary opinions are found to be equal, independent of $d$ or of the opinion's position on the spectrum (this does not hold when bias is present).

\subsection{Diffusion of opinions}
\label{subsec:fully-diff}
Consider a fully mixed population and unbiased interactions. Let $n_q(t)$ be the popularity of opinion $q$ at time step $t$. By design, at each time step there is exactly one interaction between two compatible agents. If this interaction involves the opinion $q$, $n_q$ will be changed by $+1$ or $-1$ with equal probability. If neither agent has opinion $q$, $n_q$ is not changed. Therefore:
\begin{equation}
 n_q(t) = n_q(0) + \sum_{i=1}^t s_i,
\end{equation}
where $s_i$ can be 1,0, or $-1$. Unlike the basic random walk, here the step size can be zero. Define
\begin{equation}
 \delta n_q \equiv n_q(t)-n_q(0)=\sum_{i=1}^t s_i.
\end{equation}
The diffusion law, derived from the regular one-dimensional random walk, states that the variance of position is linear in time, where the constant of proportionality is twice the diffusion coefficient \cite{Sornette2006}. To arrive at a similar relation, consider the variance of $n_q$:
\begin{equation}
 \left\langle (\delta n_q)^2\right\rangle = \left\langle \sum_{i=1}^t \sum_{j=1}^t s_i s_j \right\rangle
= \sum_{i=1}^t \sum_{j=1}^t \left\langle s_i s_j \right\rangle,
\end{equation}
where $\langle \cdots \rangle$ indicates an ensemble average (expectation value). Since $s_i$ and $s_j$ are independent random variables, their covariance $\left\langle s_i s_j \right\rangle$ is $\left\langle s_i^2 \right\rangle \delta_{ij}$. Then all cross terms vanish, yielding
\begin{equation}
 \left\langle (\delta n_q)^2 \right\rangle = \sum_{i=1}^t \left\langle s_i^2 \right\rangle = t \left\langle s_1^2 \right\rangle.
\end{equation}
The last step follows from the independence of successive steps. Therefore this process also obeys the diffusion equation with the diffusion coefficient $D=\left\langle s_1^2 \right\rangle/2$.

The next step is to calculate $\left\langle s_1^2 \right\rangle$, the expectation value of the change of popularity $n_q$ in one step. The value of $s_1^2$ is 1 if the interaction involves opinion $q$, and 0 otherwise. Therefore $\left\langle s_1^2 \right\rangle$ is the same as the probability that an interaction involves $q$. This value generally depends on $q$, on the number of opinions $Q$ and on the confidence interval $d$.

Define $p(q; Q,d)$ to be the number of opinions that are compatible with a given opinion $q$. It is given by
\begin{equation}
 p(q; Q,d) = \min(d, q-1) + \min(d,Q-q).
\end{equation}
At every time step a compatible pair of opinions are selected for interaction. If $q$ is one of these choices, there are $p(q;Q,d)$ second choices. Therefore:
\begin{equation}
 \left\langle s_1^2 \right\rangle =\frac{p(q;Q,d)}{\sum_{q=1}^Q p(q;Q,d)},
\end{equation}
where the denominator is the number of all possible interactions, which is given in section~\ref{subsec:fully-eqbtime} as $d(2Q-d-1)/2$. With this substitution the diffusion coefficient is obtained as:
\begin{equation}
\label{eq:diffcoef}
 D_{q,Q,d} = \frac{\min(d,q-1)+\min(d,Q-q)}{d(2Q-d-1)}.
\end{equation}

For given $Q$ and $d$, the diffusion coefficient increases linearly with the number of compatible opinions. Extreme opinions diffuse slowest because there are only $d$ opinions compatible with them. In contrast, middle opinions diffuse fastest as there are $2d$ opinions compatible with them.

The result~(\ref{eq:diffcoef}) indicates that, keeping $q$ and $Q$ the same, a larger value of the confidence interval $d$ leads to faster diffusion for opinion $q$. If the number of opinions $Q$ is increased while keeping $q$ and $d$ fixed, the diffusion of $q$ slows down because there are more interactions which do not involve $q$.

\section{Identical agents on networks}
\label{sec:network}
In fully mixed systems (homogeneous, completely connected), any agent can interact with any other agent directly. However, such complete interaction networks are unrealistic for all but the simplest groups. In general, one can imagine each agent occupying a node in a graph and communicating with its neighbors only. Hierarchies can be represented by directed graphs where influence goes in one direction only.

Scale-free networks or small-world networks are commonly used to simulate large communities~\cite{WattsStrogatz1998,AlbertBarabasi2002,Newman2003}. Also, networks extracted from real-life relationships can be used~\cite{CointetRoth2007}. 

If an agent represents a geographic area, a lattice of agents can be used to investigate spatial distribution of opinions and to simulate the spread (or decline) of a specific choice over a geographic region.

\subsection{Absorbing states and blocking}
\label{subsec:network-absorbing}
\begin{figure}
\includegraphics[scale=0.75]{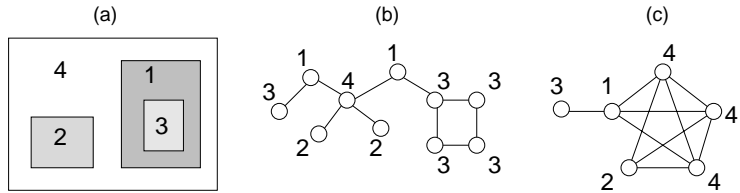}
\caption{\label{fig:endstatediag}Possible absorbing states with $Q=4$ opinions and confidence interval $d=1$. (a) Static domains on a 2D lattice, labeled with opinions. Agents (lattice sites) with compatible opinions do not have a direct connection. For example, agents with opinion 1 and agents with opinion 2 are separated by agents with opinion 4. (b) and (c) show some static states on different interaction networks. The circles are the agents and the labels indicate opinions.}
\end{figure}

In the fully-mixed system, if the opinion $q$ is a stationary opinion in an absorbing state, the popularities of all opinions compatible with $q$ must be zero (Section~\ref{sec:fully}). However, compatible opinions can coexist if the interaction network is not a complete graph. Two agents carrying compatible opinions may not be directly connected with each other, but to other agents with incompatible opinions. Figure~\ref{fig:endstatediag} shows some examples of such static cases. Such arrangements would not be possible on fully connected interactions. The scarcity of connections isolates compatible opinions, creating pluralistic absorbing states.

This \emph{blocking effect} can be observed in a wide variety of interaction networks, as long as they are not complete graphs. Even if two agents are not directly connected, they can influence each other by means of other agents located between them. Therefore, an interaction becomes more likely and blocking less likely if, on average, there are only a few agents between any given pair. In other words, the strength of the blocking effect depends on the average distance between the nodes of the underlying graph~\cite{West2001}.
 
The coexistence of compatible opinions depends not only on the interaction network, but also on the confidence interval. Figure~\ref{fig:endstatediag} illustrates how all opinions can coexist with $Q=4$, $d=1$. If, instead, the confidence bound $d$ is taken to be 2, such perfect pluralism is impossible for any network. This is because opinions 2 and 3 are compatible with all opinions, and their interactions cannot be blocked.

In general, if $Q-d \leq 1+d$ holds, then opinions labelled $Q-d$ through $1+d$ are connected to every opinion. In that case, any absorbing state can have at most $2(Q-d-1)$ stationary opinions.

\section{Discussion and conclusions}
\label{sec:concl}

This study presents a new bounded-confidence model for discrete opinions. The model addresses situations where a compromise is not possible; in any interaction one of the agents, selected randomly, converts the other to its opinion. Similar models exist, but as far as we know, this is the only model that combines arbitrary number of discrete opinions, bounded confidence, and lack of compromise. 

If the population is fully-mixed and if agents are equally likely to convert each other, the model is equivalent to the multi-player gambler's ruin problem. Using this equivalence it is shown that the system always ends up in an absorbing state. The time (number of opinion exchanges) required to reach that state scales as $N^2$ and increases monotonically with the confidence bound $d$. The distribution of absorption time is found to agree very well with the Generalized Extreme Value distribution.

In fully mixed systems with unbiased interactions, the spread of the popularity of an opinion obeys the diffusion equation. The diffusion coefficient of an opinion is proportional to the number of opinions compatible with that opinion.

When the network of individuals is a complete graph, an absorbing state consists of at least one and at most $ \lfloor(Q-1)/(d+1)\rfloor + 1$ surviving opinions. The probability of consensus, for a given $Q$, decreases with the confidence bound $d$.

When the network of individuals is not a complete graph, a given opinion may become stationary even if compatible opinions survive. This is possible if the path between compatible agents is blocked by incompatible agents.

The model is not universal; like all bounded-confidence models, it cannot be applied to preferences that cannot be ordered. In particular, it cannot be used to decide between different items.

More importantly, the assumption of non-compromise is not applicable to every opinion exchange. Consider a certain statement where the opinion value represents the level of agreement with it. A pair of individuals, arguing rationally, may agree on a mid-level opinion. Also, politics and diplomacy at every level (from personal to national) is a series of compromises.

The presentation here is deliberately limited to the most basic properties of the model. Even though this choice makes the model unrealistic in many aspects (see below), we believe it best to study the consequences of basic model, and add more interesting complications later.

The agents are assumed to be unrealistically identical. In reality, people have different levels of tolerance $d$. Also, people who are more familiar with a subject can distinguish a wider range of options (e.g., a coffee aficionado may recognize many brands with different qualities and price tags). Future work may involve heterogeneous populations with individualized $d$ and $Q$ values.

It is commonly assumed that extremists are also intolerant. In terms of the model used here, this assumption means that tolerance parameter $d$ is small for opinions 1 and $Q$ and large for central opinions. So, the tolerance parameter could depend not only on the individual, but also on the opinion of the individual. Whether this assumption is true or not requires evidence from social research. Some earlier studies~\cite{GargiuloMazzoni2008} have investigated the consequences of this assumption. The basic non-compromise model, as presented here, assumes that $d$ is constant and uniform. Intolerance of extremism can be built into the model simply by making $d$ a function of agent properties.

The basic model uses mindless agents who do not keep track of past interactions. However, history of interactions is relevant for opinion interchanges in real life. For example, the conversion probability to an opinion may be proportional to how many times an agent has encountered a follower of that opinion. This, too, can be added to the model in later studies, resulting in different absorbing state probabilities.

This study assumes that the number of opinions is given, and constant. However, in real life, new options (coffee brands, political parties) appear all the time, diversifying the choice portfolio. The basic model can be extended to create new opinions, and the consequences of choice overload can be simulated.

Real communities do not form a complete graph. On the macroscopic level, the interaction network of agents may be changed, and properties of diffusion and absorbing states can be compared with the fully-mixed case. Some obvious modifications include higher dimensional lattices, random networks, small-world networks or scale-free networks.

The model can be further generalized by constructing an \emph{opinion network}, where opinions are the vertices, and two opinions A and B are connected with an edge if they are deemed ``close'', so that one agent with opinion A would convert to opinion B after the interaction (or vice versa). The opinion network approach is more general than expressing bounded confidence as the difference between numerical values, as it allows the construction of other interesting opinion relations (e.g., a star graph, where opinions are incompatible with all other opinions except for the one at the hub). The model used here is a special case, where the opinion network is a path graph from 1 to $Q$ if the tolerance $d$ is 1, and it is a complete graph if $d=Q-1$. 

\section*{Acknowledgements}
I am thankful to T\"urker B\i y\i ko\u{g}lu, Gunnar Pruessner, \.{I}smail Karakurt,
Mehmet Ayd\i n, Avadis Hac\i nl\i yan, G\"ulden G\"uven\c{c} for useful discussions, to Meral \"Ozt\"urk for proofreading, and to the reviewers and editors for their helpful comments.

\bibliographystyle{ws-acs}
\bibliography{references}
\end{document}